\pdfoutput=1

\documentclass[10pt]{article}

\usepackage{authblk}                
\usepackage[square,numbers]{natbib} 

\usepackage{amsmath,amssymb}                

\usepackage{listings}    
\usepackage{graphicx}    
\usepackage{subcaption}  
\usepackage{algorithm}   
\usepackage{algorithmic} 
\usepackage{booktabs}    
\usepackage{hyperref}    
\usepackage{xurl}        
\usepackage[htt]{hyphenat} 
\usepackage{microtype} 
\DisableLigatures{}    

\usepackage{caption}
\usepackage{setspace}
\usepackage{multirow}
\usepackage{enumerate}
\usepackage{pdflscape}
\usepackage{pgfplots} 

\usepackage{xcolor}
\definecolor{codegreen}{rgb}{0.25,0.5,0.35}
\definecolor{codegray}{rgb}{0.5,0.5,0.5}
\definecolor{codepurple}{rgb}{0.6,0,0}
\definecolor{backcolour}{rgb}{0.95,0.95,0.92}
\definecolor{colorstring}{rgb}{0.5,0,0.35}
\definecolor{rltred}{rgb}{0.5,0,0}
\definecolor{rltgreen}{rgb}{0,0.5,0}
\definecolor{rltblue}{rgb}{0,0,0.5}
\definecolor{DarkGreen}{rgb}{0.00,0.60,0.00}
\definecolor{ScarletRed}{rgb}{0.80,0.00,0.00}
\definecolor{blizzardblue}{rgb}{0.67, 0.9, 0.93}
\definecolor{green-yellow}{rgb}{0.68, 1.0, 0.18}
\definecolor{dkgreen}{rgb}{0,0.6,0}
\definecolor{gray}{rgb}{0.5,0.5,0.5}
\definecolor{mauve}{rgb}{0.58,0,0.82}
\definecolor{lightgrey}{rgb}{0.90,0.90,0.90}
\definecolor{grey}{gray}{0.75}
\definecolor{light-gray}{gray}{0.80}

\lstdefinestyle{mystyle}{
    escapechar=©, 
    language=Python,
	backgroundcolor=\color{backcolour},
    basicstyle=\footnotesize\ttfamily,
   	identifierstyle=\footnotesize\ttfamily,
	commentstyle=\color{codegreen},
	keywordstyle=\color{colorstring}\bfseries,
	morekeywords={OR, AND},
	numberstyle=\ttfamily\color{codegray},
	stringstyle=\ttfamily\color{DarkGreen},
	breakatwhitespace=false,
	breaklines=true,
	captionpos=b,
	keepspaces=true,
	numbers=left, 
	numbersep=2pt,
	showspaces=false,
	showstringspaces=false,
	showtabs=false,
	tabsize=2
}
\usepackage{braket}		
\usepackage{mathtools}	

\usepackage{pifont} 

\usepackage{xspace}

\usepackage{tcolorbox}
\newtcolorbox{resultsbox}[1][]
{
	colframe=gray!100,
	colback=white!100,
	fonttitle=\bfseries,
	coltitle=white,
	title=#1
}

\newtcolorbox{mybox}[2][]{
	colframe=#2!30!black,
	colback=#2!3,,
	coltitle=white,
	fonttitle=\bfseries,
	title=#1,
	boxrule=0.8pt,
	arc=1mm,
	left=1pt,
	right=2pt,
	top=1pt,
	bottom=0pt
}

\newenvironment{examples}[1][]{
	\begin{mybox}[#1]{black}
		\small
}{
	\end{mybox}
}

\usepackage{ifthen}
\newboolean{showcomments}
\setboolean{showcomments}{true} 

\ifthenelse{\boolean{showcomments}}{
	\newcommand{\nbc}[3]{
		{\colorbox{#3}{\bfseries\sffamily\scriptsize\textcolor{white}{#1}}}
		{\textcolor{#3}{\sf\small$\langle$\textit{#2}$\rangle$}}}
	
}{
	\newcommand{\nbc}[3]{}

}

\newcommand{\squeezeup}{\vspace{-4mm}}
\newcommand{\secsqueezeup}{\vspace{-2pt}}

\usepackage{geometry}
\title{
Managing Uncertainty \\in LLM-based Multi-Agent System Operation

}

\author[1]{Man Zhang}
\author[1]{Tao Yue\thanks{Corresponding author}}
\author[2]{Yihua He}

\affil[1]{Beihang University}
\affil[2]{Maternal-Fetal Medicine Center in Fetal Heart Disease, Capital Medical University, Beijing Anzhen Hospital}
\affil[ ]{\texttt{\{manzhang, yuetao\}@buaa.edu.cn}, \texttt{heyihuaecho@hotmail.com}}

\date{}

\begin{document}

\maketitle

\begin{abstract}
Applying LLM-based multi-agent software systems in safety-critical domains such as lifespan echocardiography introduces system-level risks that cannot be addressed by improving model accuracy alone. 
During system operation, beyond individual LLM behavior, uncertainty propagates through agent coordination, data pipelines, human-in-the-loop interaction, and runtime control logic. Yet existing work largely treats uncertainty at the model level rather than as a first-class software engineering concern.
This paper approaches uncertainty from both system-level and runtime perspectives. 
We first differentiate epistemological and ontological uncertainties in the context of LLM-based multi-agent software system operation. 
Building on this foundation, we propose a lifecycle-based uncertainty management framework comprising four mechanisms: representation, identification, evolution, and adaptation. The uncertainty lifecycle governs how uncertainties emerge, transform, and are mitigated across architectural layers and execution phases, enabling structured runtime governance and controlled adaptation.
We demonstrate the feasibility of the framework using a real-world LLM-based multi-agent echocardiographic software system developed in clinical collaboration, showing improved reliability and diagnosability in diagnostic reasoning. The proposed approach generalizes to other safety-critical LLM-based multi-agent software systems, supporting principled operational control and runtime assurance beyond model-centric methods.
\end{abstract}


\section{Introduction}
\label{sec:introduction}
Large Language Model (LLM)-based multi-agent software systems represent a paradigm shift where LLMs serve as the central reasoning core for processing complex, multi-modal data~\cite{hou2024harnessing,chao2025evaluating}. A prominent example is the development of LLM-basesd systems for lifespan echocardiography which integrate LLMs with cardiac imaging and clinical records to provide context-aware diagnostic reasoning that adapts to unique physiological changes from infancy through old age.
Such applications leverage foundation models (e.g., DeepSeek~\cite{liu2024deepseek} and GPT-4~\cite{achiam2023gpt}) to interpret high-dimensional data, but introduce systemic complexities that traditional architectures were not designed to handle.

In high-stakes domains such as disease diagnosis, managing uncertainties is fundamental for clinical safety~\cite{simpkin2016tolerating}, long before LLMs became available. 
LLM-based multi-agent systems are uniquely susceptible to hallucinations, where plausible but factually incorrect diagnostic justifications are generated, as well as inherent uncertainties in the operating environment and sensor noise \cite{xia2025survey}. Furthermore, out-of-distribution clinical phenotypes (e.g., heterotaxy with situs ambiguous) or shifts in the deployment infrastructure (e.g., microservice scaling) create non-deterministic behaviours. Without rigorous operational bounding, these fluctuations can lead to potentially catastrophic in time-critical cardiac assessments.

Existing uncertainty-aware software engineering research~\cite{UncerTum2019, han2023uncertainty, zhang2019uncertainty, troya2021uncertainty, xu2022uncertainty, bertoa2020incorporating, catak2021prediction} provides a strong foundation, but it must be updated to address the distinct challenges of LLM-based multi-agent software systems. Established methodologies for uncertainty-aware specification and analysis~\cite{zhang2018URUCM, shin2021uncertainty, zhang2020uncertainty, han2023uncertainty}, uncertainty modeling~\cite{zhang2019uncertainty, bandyszak2020orthogonal, bertoa2018expressing, makelburg2025surveying}, uncertainty-aware testing~\cite{zhang2019UncerTest, zhang2017uncertainty, camilli2021uncertainty}, and uncertainty-aware simulation~\cite{jezequel2023uncertainty} were largely proposed for systems with well-defined operational boundaries and explicit logic paths. 
With the rise of LLMs, a growing body of work has emerged to study uncertainties inherent to the models themselves, primarily focusing on uncertainty quantification for token generation and confidence calibration~\cite{liu2025survey, shorinwa2025survey}, scenario-independent uncertainty estimation~\cite{wen2025scenario} to disentangle semantic truth from lexical noise, the detection of internal hallucinations through self-probing and consistency check~\cite{manakul2023selfcheckgpt}, etc.
However, there is currently a lack of systematic research that addresses uncertainty from the perspective of LLM-based multi-agent software systems, where the LLM serves as a reasoning core integrated with possibly multi-modal data, dynamic microservices, and human-in-the-loop (HITL) workflows. In domains like lifespan echocardiography, managing uncertainty requires concerning both model-level metrics, system-wide infrastructure, and the human-machine interface (HMI). This multi-layered perspective ensures that risks are not only detected within the computational core but also effectively communicated and mitigated through collaborative workflows between the autonomous agents and human experts. 

In this paper, we aim to systematically examine the diverse types of uncertainty inherent in LLM-based multi-agent software systems, illustrating our findings through a real-world industrial case study of an LLM-based echocardiographic platform. Based on these understandings, we propose a strategic vision for the comprehensive management of uncertainty in such systems. We specifically draw attention to the recently published Precise Semantics for Uncertainty Modeling (PSUM) international standard \cite{omg2024psum} standardized at the Object Management Group (OMG), which was built on years of effort of the community, especially the uncertainty modeling conceptual model: U-Model~\cite{zhang2016understanding}. 
We argue that while PSUM was defined prior to the widespread adoption of LLMs, its belief-centered approach remains highly relevant for characterizing various types of uncertainties in LLM-based multi-agent software systems. Our long term is to leverage PSUM to develop a systematic infrastructure that enables the active management of diverse uncertainties throughout their lifecycles.

Specifically, PSUM~\cite{omg2024psum} aims to make uncertainty explicit within a defined model by attaching it to \textit{BeliefStatement}s (tangible representations of a \textit{BeliefAgent}'s knowledge) through \textit{UncertaintyTopic}. The PSUM metamodel characterizes \textit{Belief}, \textit{IndeterminacySource}, \textit{Uncertainty}, and define relationships among them and associate them to \textit{Evidence} and \textit{Risk}. Especially, rather than treating uncertainty purely as a numerical attribute, PSUM models uncertainty as a belief-centered construct grounded in identifiable causes of indeterminacy. PSUM also provides the way of quantifying uncertainty and other \textit{MeasurableElement}, such that numerical or symbolic values can be assigned to features such as \textit{Accuracy} or \textit{Precision}. 

We envision an uncertainty-aware framework for managing the evolving uncertainty inherent in LLM-based multi-agent systems during their operation. Given the complexity of autonomous agents, heterogeneous data sources, and intricate inter-agent interactions, our framework adopts a systematic \textit{uncertainty lifecycle} approach, which transitions through states such as \textit{Detected}, \textit{Characterized}, \textit{Mitigated}, and \textit{Resolved}. This lifecycle is governed by four core mechanisms (\textit{Representation}, \textit{Identification}, \textit{Evolution}, and \textit{Adaptation}). By aligning these mechanisms, in a long term, we aim to achieve three critical objectives: the explicit representation of uncertainty, its continuous and progressive handling, and adaptable system operation under uncertainty. Central to our vision is the elevation of human actors' roles to uncertainty-aware agents who participate selectively and accountably within the uncertainty lifecycle, to ensure that human-machine collaboration is both transparent and grounded in the system's current state of belief.

The rest of this paper is organized as follows: Section~\ref{sec:classification} provides a systematic classification of uncertainties within the context of LLM-based multi-agent systems. Section~\ref{sec:framework} introduces our vision for an uncertainty management framework, detailing its core mechanisms and the proposed uncertainty lifecycle. Finally, Section~\ref{sec:conclusions} concludes the paper and outlines directions for future research.

\secsqueezeup
\section{Understanding Uncertainties}\label{sec:classification}

\begin{figure}[t]
	\centering
	\includegraphics[width=.6\linewidth]{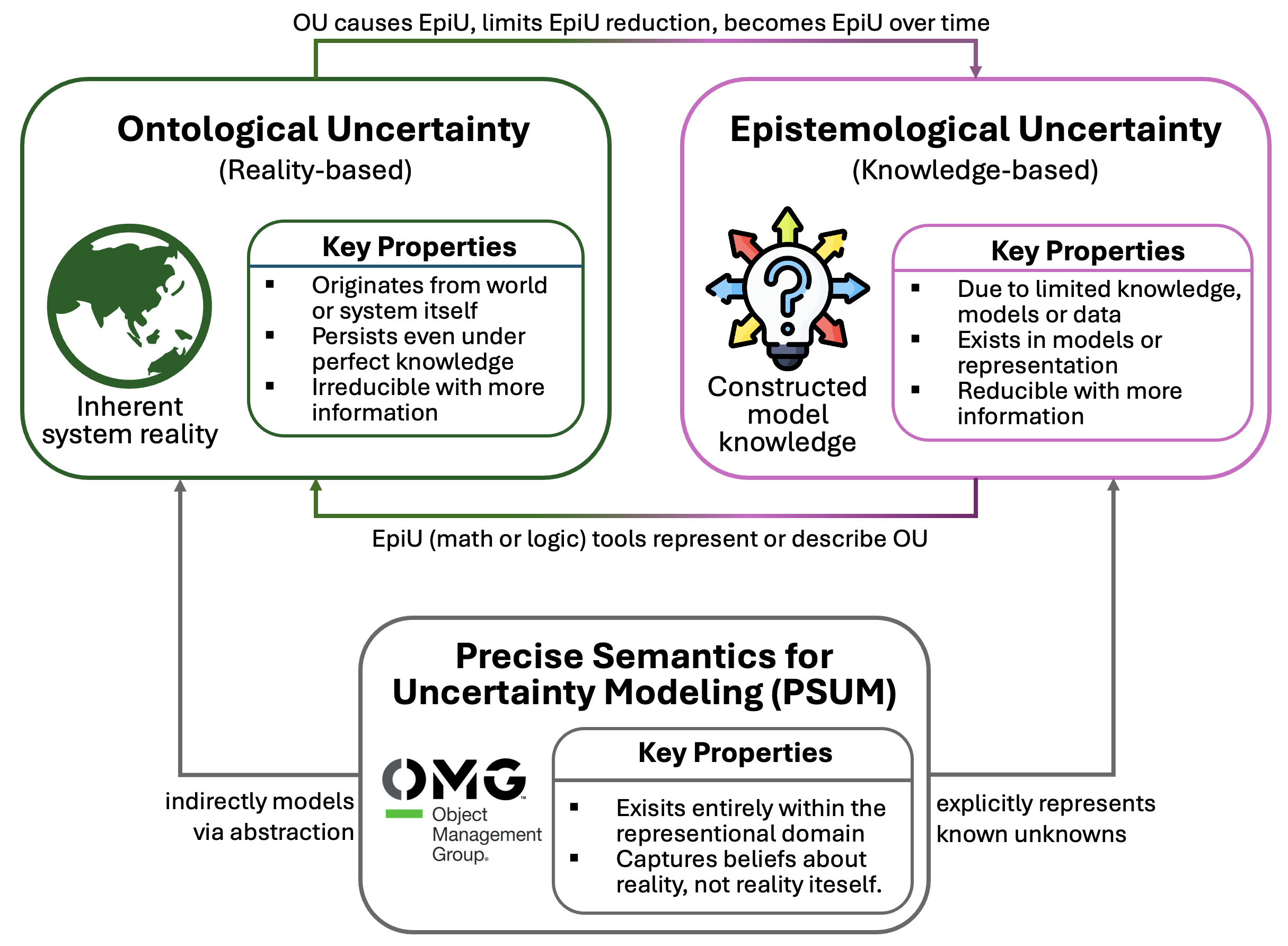}
	\caption{Overview of Epistemological Uncertainty, Ontological Uncertainty, PSUM, and their Relations.}
	\label{fig:epi-ont-psum}
	\squeezeup
\end{figure}

The inherent complexity of LLM-based multi-agent software systems (e.g., LLM-based systems for lifespan echocardiography) necessitates a careful understanding of the diverse forms of uncertainty. By identifying these distinct types, it becomes possible to move beyond treating the system as an opaque ``black box'' and instead develop targeted methodologies to manage these uncertainties, which span ontological origins, learning dynamics, model behaviour, and decision-making impact. 
 
We classify such uncertainties into: epistemological uncertainty (EpiU) from ontological uncertainty (OU), as illustrated in Figure~\ref{fig:epi-ont-psum}. 
\textit{Epistemological uncertainty} arises from limited, imperfect, or abstracted knowledge about a system, its data, or its models, and has long been recognized as a fundamental source of uncertainty in modelling and decision-making~\cite{walker2003defining, der2009aleatory}. It is in principle reducible with better models, data, or reasoning. This type of uncertainty is essentially about what we know and how we represent what we know, not in the system itself. Hence, it is \textit{knowledge-based}.
\textit{Ontological uncertainty} arises from inherent properties of the system or environment, is independent of our knowledge, and is irreducible even with complete information, and has been discussed extensively in philosophical and systems literature~\cite{lane2005ontological, klinke2024theory}. It is in principle about uncertainty in the system (hence \textit{reality-based}), not in our description of the system. 
The reality's complexity forces the observers to be ignorant (``OU causes EpiU''); one cannot know a result more precisely than the source's inherent indeterminacy allows, which effectively limits the reduction of uncertainty (``OU limits EpiU reduction''). Over time, ontological events may transit into EpiU as they move from potential (future) occurrences to (already-occurred) historical facts that may be poorly recorded or understood (``OU becomes EpiU over time''). Understanding the relations helps to systematically management uncertainties.

%


\subsection{Epistemological Uncertainty}
\label{subsec:EpiU}
\begin{figure}
	\centering
	\includegraphics[width=.7\linewidth]{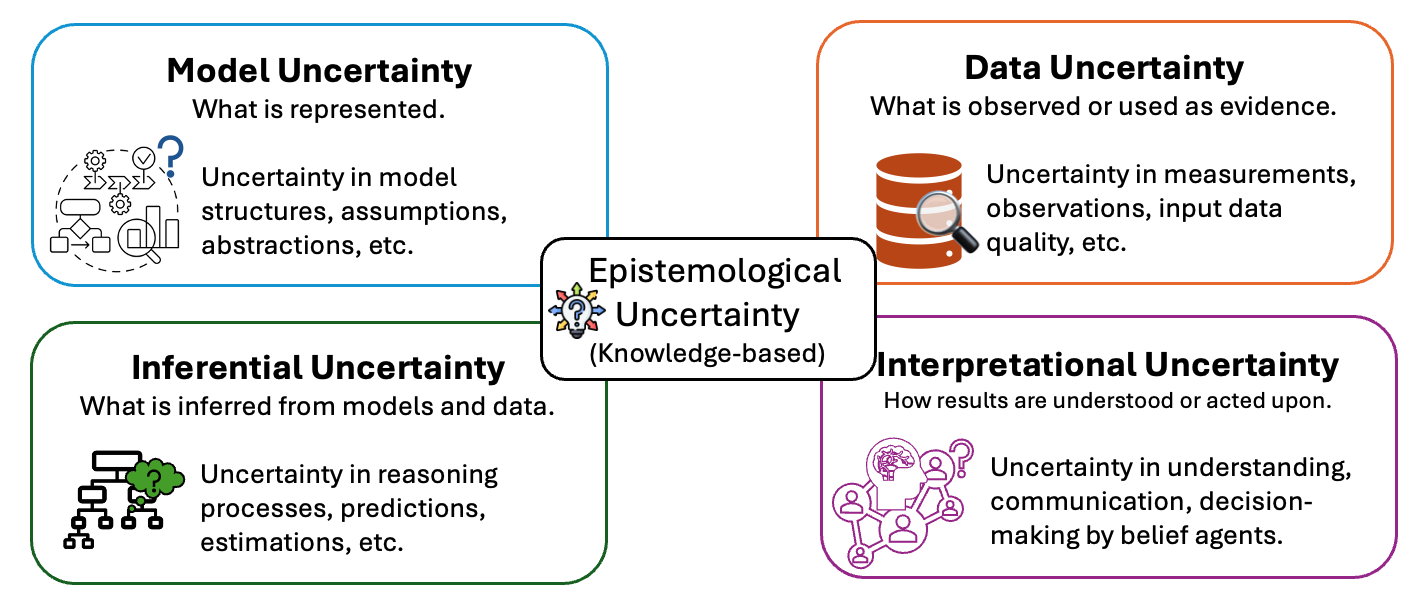}
	\caption{Epistemological Uncertainty} 
	\label{fig:epiTaxonomy}
\end{figure}

As shown in Figure~\ref{fig:epiTaxonomy}, the four epistemological uncertainty types are conceptually orthogonal. 


\subsubsection{Model Uncertainty}

%
%
%
%

In general, a model is an abstract representation of phenomenon or knowledge domains constructed to support reasoning, prediction, explanation, or decision-making, by omitting detail and idealizing reality. 
Since every model is built via abstraction, such as selecting scope, granularity, assumptions, and representational forms, which inevitably introduces epistemological uncertainty. This also reflects the famous statement by George Box: ``all models are wrong, but some are useful''~\cite{box1976science}. 
Taking LLMs as example, they can never perfectly replicate the real world, but they are incredibly useful by finding patterns that help experts make better decisions such as echocardiographic diagnosis.

We distinguish four types of model uncertainty: structural, behavioural, parameter, and semantic, which all arise from abstraction choices made during modelling, and together shape the adequacy of the model representation. In contrast, applicability uncertainty concerns whether these abstractions remain appropriate for a given application context. 


\emph{Structural uncertainty} in operation concerns the dynamic selection and orchestration of the system's reasoning components and their relationships. For instance, as described in ~\cite{wang2025survey}, adaptive planning architectures allow these systems to adjust decision-making frameworks at runtime based on task complexity. In such architectures, the planning layer continuously updates clinical strategies, prioritizes urgent cases, and refines previous decisions as findings emerge, to enable specialized responses to diverse patient needs, adapt to varying task complexities and fluctuating computational resources, and integrate real-time evidence, which inevitably introduces structural uncertainty during system operation. 


\emph{Behavioural uncertainty} in operation concerns the non-deterministic variability in a system’s internal execution logic and process-level transitions. As evidenced in adaptive medical agent frameworks~\cite{wang2025survey}, this uncertainty is driven by the stochastic nature of the reasoning and planning layers, where the system may follow divergent reasoning traces or trigger variable self-correction loops for identical inputs. Notably, behavioural uncertainty frequently serves as a primary driver of inferential uncertainty (see Section~\ref{subsub::inferentialU}); the instability in the derivation process (how the conclusion is reached) directly leads to the doubt in the final prediction or decision (what is concluded). 

\emph{Parameter uncertainty} in operation concerns uncertainty in numerical or symbolic values that control model behaviour at runtime. This goes beyond static weights of a trained model and include configurable thresholds for task complexity, priority scoring within the adaptive planning layer of an LLM-based multi-agent systems. 

\emph{Semantic uncertainty} concerns about the real-world meaning and interpretation of model constructs, such as uncertainty in the correspondence between model elements and domain concepts, uncertainty about how learned internal representations in models correspond to intended domain semantics. 
This uncertainty manifests in multi-agent orchestration, where specialized agent roles may lack a shared semantic grounding, leading to misinterpretations of perception data and misaligned reasoning traces.

\emph{Applicability uncertainty} concerns if a model’s internal abstractions and foundational assumptions remain valid when the system is applied beyond its intended operational envelope. 

\begin{examples}[Model Uncertainty - Examples]
	\textbf{Structural Uncertainty.} The system's adaptive planning layer decides at runtime whether to orchestrate a single-agent diagnostic trace for a routine case or a collaborative expert group (consisting of specialized pediatric, surgical, and hemodynamic agents) to interpret complex congenital heart defects.
	
	\textbf{Behavioural Uncertainty.} Inconsistent task sequencing or reasoning paths in an echocardiography agent can lead to poorly calibrated confidence scores, as the underlying derivation procedure lacks a stable, reproducible execution logic.
	
	\textbf{Parameter uncertainty.} An LLM-based agent may assign varying weights to clinical markers (e.g., prioritizing certain echo measurements over others) based on the past cases. Any operational miscalibration in these dynamic weights can lead to biased diagnostic inference.
	
	\textbf{Semantic uncertainty.} A general triage agent might interpret a heart rate of 140 bpm as a``critical'' emergency, while a pediatric agent interprets the same value as "normal" for a neonate. This discrepancy in meaning can lead to misaligned reasoning traces and incorrect clinical planning.
	
	\textbf{Applicability uncertainty.} Applying the system trained on high-resolution adult scans to handheld pediatric POCUS introduces applicability uncertainty, as the model’s learned abstractions regarding image quality may not generalize. 
\end{examples}

\subsubsection{Data Uncertainty}

Data uncertainty in operation concerns the quality, reliability, and representativeness of the observations or datasets used to execute or evolve models. This includes noise in input signals, incomplete data records, or distributional instability where the operational data significantly drifts from the system’s initial training distribution~\cite{xu2025LLM}. Within a multi-agent framework, when using uncertain data to evolve the model, data uncertainty fundamentally alters the system’s future reasoning logic and parameters; when using uncertain data solely to operate a fixed agent, the uncertainty propagates to the output without altering the underlying model representation.
We classify data uncertainty into four categories: noise uncertainty, missing data, sample bias, and distributional shift, because together they capture the fundamental ways in which data can be epistemically imperfect. 

\textit{Noise uncertainty} arises from incomplete knowledge, imperfect characterization, or the simplified modeling of perturbations affecting observed data. In LLM-based systems, this is not limited to physical sensor noise; it critically includes semantic noise (ambiguity, subjectivity, or inconsistency in how data values or labels are defined and assigned). When agents operate on such ``noisy'' inputs, the lack of a precise ground truth can trigger divergent reasoning traces or unnecessary reflection loops, as the system struggles to distinguish between meaningful signals and artifacts of imperfect data characterization.

%

%
\textit{Missing data} is caused by the absence of values for some variables or instances, which consequently restricts the understanding about the modelled phenomenon. 
%

\textit{Sampling bias} in operation refers to the epistemic limitation where the subset of information actively used or retrieved by the system is not representative of the application context. This leads to a contextual mismatch where a representative reasoning trace is impossible because the agent is operating on a skewed fragment of the available truth.
%

\textit{Distributional shift} arises when the statistical properties of the data used during model construction differ from those encountered during model application. 
%

\begin{examples}[Data Uncertainty - Examples]
	\textbf{Noise uncertainty.} In LLM-based lifespan echocardiography, noise uncertainty arises from unmodeled physical perturbations (e.g., patient respiratory cycles) and semantic inconsistencies (e.g., subjective expert labeling), which introduce variability into the diagnostic data. 

	\textbf{Missing data.} The absence of specific Doppler flow measurements or obscured anatomical views in neonatal scans prevents an agent from satisfying the logical preconditions required to transition from raw observation to a finalized clinical plan, often forcing the system to request human clarification.
	
	\textbf{Sampling bias.} The system introduces uncertainty through the biased selection of evidence from its experience base or the incomplete sampling of patient features during multi-agent handoffs, i.e., the loss of critical information when one agent summarizes or filters a patient's data before passing it to the next agent. 
	
	\textbf{Distributional shift.} If a medical facility adopts a new ultrasound device that produces measurements with a higher mean value than those in the LLM's training set, in operation, the triage agent may over-report ``abnormal'' findings because its internal thresholds are no longer calibrated to the new data distribution, potentially triggering unnecessary and costly collaborative expert group sessions for healthy patients.
	
\end{examples}

\subsubsection{Inferential Uncertainty}\label{subsub::inferentialU}
Inferential uncertainty is about uncertainty introduced during the process of deriving conclusions, making predictions or decisions from a given model and data. Such uncertainties reflect limitations in inference, estimation, or decision procedures rather than in the model structure or data themselves. Inferential uncertainty can be commonly classified into prediction uncertainty and calibration uncertainty.

\textit{Prediction uncertainty} refers to the quantified doubt in the model’s generated output for a specific input set. 
%
\textit{Calibration uncertainty} arises when a confidence score becomes an unreliable reflection of reality, often due to biased or outdated factors such as a stale dataset or a significant change in the operating context. 


\begin{examples}[Inferential Uncertainty - Examples]
	\textbf{Prediction uncertainty.} When evaluating a neonatal echocardiogram with missing Doppler flows, the system may output a diagnosis of ``suspected Patent Ductus Arteriosus (PDA)'' with a belief score of 0.72. This prediction uncertainty signals to the clinician that the lack of complete physiological data prevented the system from reaching the high-confidence threshold required for a definitive diagnostic claim.
	
	\textbf{Calibration uncertainty.} If a model trained on adult data provides a 95\% confidence score for a pediatric diagnosis, that score may be poorly calibrated and thus an unreliable reflection of true accuracy. This requires a meta-level evaluation of whether the assigned probability can be trusted as a valid representation of the system's diagnostic competence in a specific clinical context. 
\end{examples}

\subsubsection{Interpretational Uncertainty}
It concerns uncertainty about how results should be understood, communicated, or acted upon by humans, organizations and automated agents. It is not uncertainty in the model, data, inference outcomes or confidence values, and comes after the inference is complete. 
Interpretational uncertainty can be decomposed into \textit{semantic ambiguity}, \textit{explanation uncertainty}, and \textit{interpretation variance} across belief agents, capturing uncertainty in meaning (e.g., vague labels), justification (e.g., why a specific result is produced), and agent-dependent understanding of results (e.g., different agents interpreting results differently). 

More specifically, \textit{semantic ambiguity} is about the meaning of model outputs, concepts, or labels, arising when their semantics are underspecified, overloaded, or context-dependent. 
%
%
\textit{Explanation uncertainty} is about the adequacy, faithfulness, or completeness of explanations provided for model results, regardless of whether the explanations are causal or non-causal.  
%
%
\textit{Interpretation variance} arises when different belief agents draw different conclusions or actions from the same result. 

\begin{examples}[Interpretational Uncertainty - Examples]
	\textbf{Semantic ambiguity.} A diagnostic output such as ``increased chamber size'' has multiple interpretations depending on the patient's age, making the clinical significance of the AI’s finding unclear without further context.
	
	\textbf{Explanation uncertainty.} While an LLM may correctly diagnose a congenital heart defect, the accompanying explanation (e.g., citing specific flow velocities) may be unfaithful to the actual data processed or incomplete in its clinical reasoning, which leaves the cardiologist uncertain as to if the LLM's conclusion is grounded in trustworthy evidence.
	
	\textbf{Interpretation variance.} An LLM and cardiologists derive different clinical actions from the same diagnostic report. 
\end{examples}

\subsection{Ontological Uncertainty}
\begin{figure}
	\centering
	\includegraphics[width=.6\linewidth]{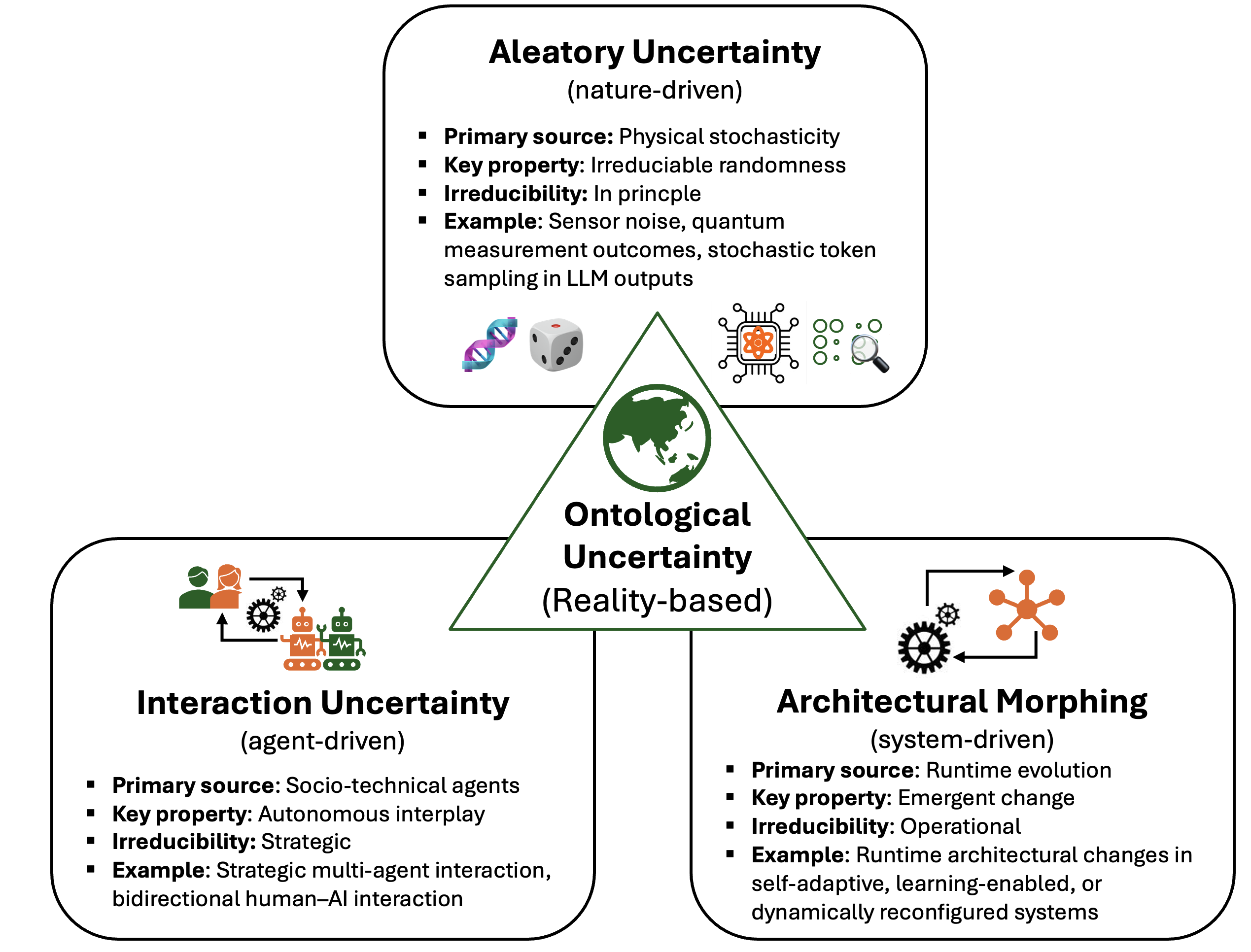}
	\caption{Ontological Uncertainty} 
	\label{fig:ontoTaxonomy}
\end{figure}

As shown in Figure~\ref{fig:ontoTaxonomy}, we classify ontological uncertainties into three categories. \textit{Aleatory uncertainty} is \textit{nature-driven}, as its irreducible randomness originates from physical stochasticity inherent in the world, such as unavoidable signal noise in ultrasound physics or environmental variability during an echocardiographic scan.
\textit{Architectural morphing} is system-driven. In the context of LLM-based multi-agent software systems,  it focuses on the underlying infrastructure of the applications, which may arise from runtime changes in the system’s own structure, such as the dynamic scaling of microservices, updates to the deployment pipeline, or the evolving internal decision logic of the LLM as it adapts to new data. 
\textit{Interaction uncertainty} is agent-driven, stemming from the decentralized and interdependent decision-making between the system and its operators. For instance, in human-in-the-loop scenarios, the mutual influences and feedback loops between the cardiologist and the AI lead to collaborative outcomes that cannot be deterministically predicted from the actions of either agent in isolation.

\subsubsection{Aleatory Uncertainty}
It refers to irreducible randomness inherent in the behaviour of a system or its operating environment. It has been intensively discussed in AI~\cite{tahir2023fairness, wang2025aleatoric}. 
The key property is that aleatory uncertainty comes from randomness in physical or computational processes; hence, even with perfect monitoring or control, outcomes remain stochastic and hence irreducible in principle or by definition.  

\subsubsection{Architectural Morphing}
With architectural morphing, we aim to describe the runtime evolution of a system’s internal structure, its configuration and composition \textit{during execution}. Its key property is \textit{emergent change} in the sense that the system’s topology or behaviour arises at runtime. Note that the adaptation logic might be pre-defined, but a specific configuration is triggered by shifting environmental configurations or evolving learning objectives. 
This is prevalent in self-adaptive systems that autonomously reconfigure their architectures such as dynamically swapping or scaling microservices~\cite{luo2022power} and in learning-enabled systems where an AI agent’s internal decision logic evolves through continuous reinforcement learning~\cite{khetarpal2022towards}. 

Architectural morphing is different from aleatory uncertainty, regarding irreducibility. Aleatory uncertainty is irreducible \textit{in principle}; but architectural morphing is irreducible operationally in the sense that, even with perfect knowledge, the runtime structure changes of a system are driven by system adaptation mechanisms, which cannot be fully resolved at design time.


\subsubsection{Interaction Uncertainty}
It originates at the socio-technical boundary where a system interacts with humans, organizations, or other AI agents, whose behaviours are strategic and uncontrollable. 
The key property of interaction uncertainty is \textit{autonomous interplay}, where outcomes result from decentralized decision-making of belief agents rather than a single centralized logic. We see three different types of interaction dynamics: 1) reciprocal feedback loops, where system actions trigger adaptive responses of agents; 2) strategic interdependence, where outcomes (e.g., in LLM-enabled multi-agent systems~\cite{chen2025static}) depend on agents reacting to each other's actions or outputs; and 3) decentralization coordination, which arises when no single agent has a global view or control of the whole system, so its overall behaviour are resulted from many local decisions, which eventually makes the system’s behaviour hard to predict or explain. 
Hence, interaction uncertainty exhibits strategic irreducibility, as system behaviour emerges from mutually adaptive decisions among agents and cannot be decomposed into independent or centrally analyzable behaviours.

\begin{examples}[Ontological Uncertainty - Examples]
	\textbf{Aleatory uncertainty.} The inherent acoustic speckle and thermal noise in ultrasound physics is aleatory uncertainty.
	
	\textbf{Architectural morphing.} It might be the dynamic scaling of microservices, or adaptive offloading of computational tasks between edge and cloud environment, or switching between adult and pediatric models at runtime. 
	
	\textbf{Interaction uncertainty.} It often stems from non-deterministic and decentralized collaboration between the cardiologist and the LLM-based agents, where the final clinical decision emerges from iterative feedback loops and competing agencies that cannot be predicted by analyzing either agent in isolation.
		
\end{examples}

\section{Uncertainty Management Framework}\label{sec:framework}

\begin{figure}[t]
	\centering
	\hspace*{-0.4cm}
		\includegraphics[width=.99\linewidth]{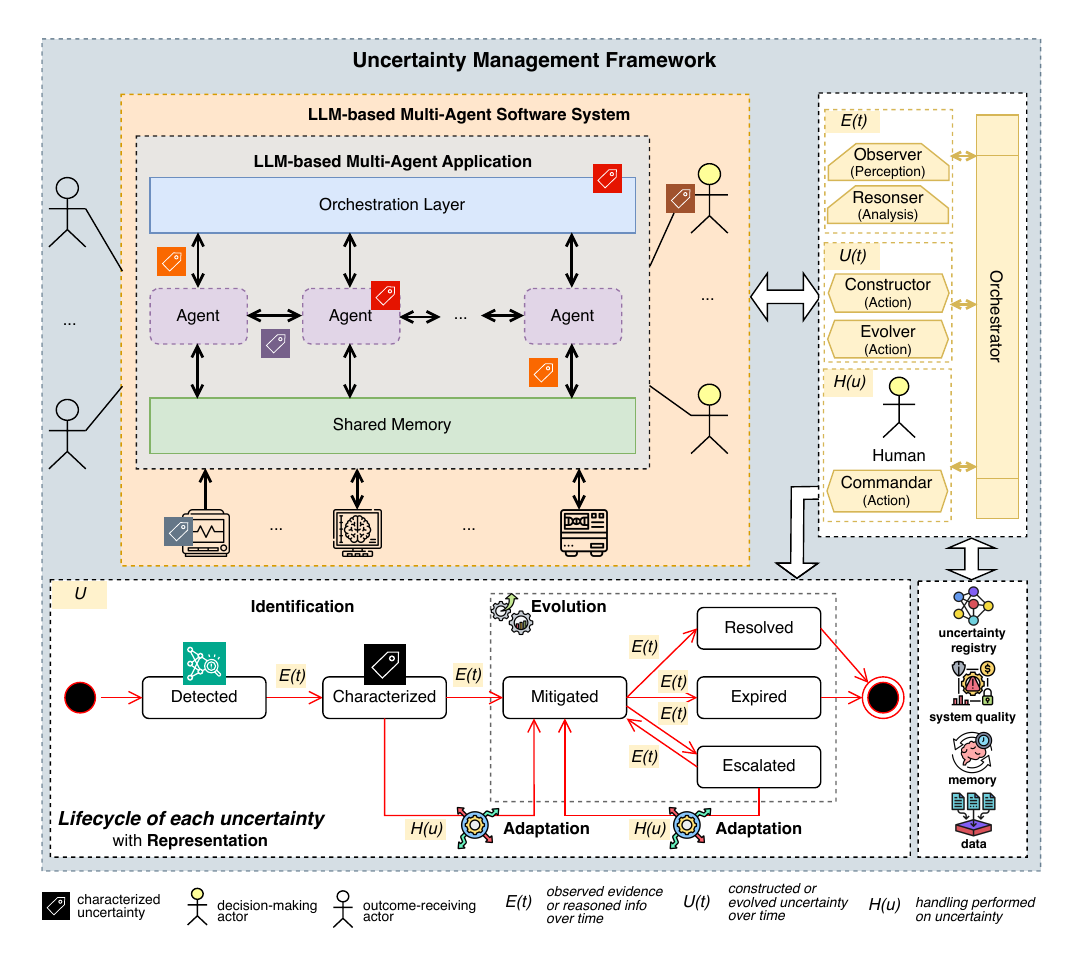}
	\vspace{-1cm}
	\caption{Overview of Uncertainty Management Framework} 
	\label{fig:overview}
	\vspace{-0.4cm}
\end{figure}

We propose an uncertainty-aware framework for managing uncertainty in LLM-based multi-agent software systems during operation. 
Such systems typically involve multiple autonomous agents, heterogeneous data sources, and complex inter-agent interactions, which subsequently introduce evolving forms of uncertainty in their operation.
To address these challenges, our framework manages uncertainty via {four} mechanisms (i.e., \textbf{Representation}, \textbf{Identification}, \textbf{Evolution} and \textbf{Adaptation}) that span the entire \textbf{uncertainty lifecycle}. 
These mechanisms are realized via role-based multi-agents (as shown in Figure~\ref{fig:overview}), where uncertainty is perceived, analyzed, evolved, and acted upon in a coordinated and policy-governed manner, aligned with three key objectives:

\emph{\textbf{Objective 1: Explicit uncertainty representation.}}
Uncertainty is manageable only if it and its propagation are explicitly represented or annotated, rather than remaining implicit.
Therefore, our first objective is to explicitly represent and characterize uncertainty, especially epistemological uncertainty, arising during system operation, such as data, models, reasoning, and agent interactions.
Note that epistemological uncertainty can often be explicitly represented and progressively reduced through additional evidence, deliberation, or verification, while ontological uncertainty may not be fully eliminable or explicitly resolvable, as we discussed in Section~\ref{sec:classification}.

\emph{\textbf{Objective 2: Continuous and progressive uncertainty handling.}}
As LLM-based agents observe new information, interact with one another, and incorporate human input, uncertainty may increase, decrease, or persist.
Managing uncertainty needs to be an evolving process rather than a one-time handling.
Our framework aims at systematically identifying both epistemological and ontological uncertainty as an LLM-based multi-agents system operates.

\emph{\textbf{Objective 3: Adaptable system operation under uncertainty.}}
To ensure that an LLM-based multi-agent system continues to function properly in the presence of uncertainty,
according to its severity and associated risk, our framework aims to safeguard system operation by adapting autonomy levels, coordination strategies, and decision-making behaviour.
This enables bounded, and risk-aware system behaviour in operation. 
Unlike approaches that assume complete or accurate operational information, an assumption that is rarely practical, our framework explicitly treats uncertainty as a first-class concern.

\vspace{-0.4cm}
\subsection{Uncertainty Lifecycle}

In the context of our uncertainty management framework, we define the lifecycle of each uncertainty \(U\) is composed of six states, i.e., \textit{Detected}, \textit{Characterized}, \textit{Mitigated}, \textit{Resolved}, \textit{Escalated}, and \textit{Expired}, governed by the identification, representation, evolution, and adaptation mechanisms.
As illustrated in Figure~\ref{fig:overview}, transitions between states are driven by time-indexed evidence \(E(t)\) and uncertainty-handling actions \(H(u)\), reflecting the dynamic nature of uncertainty during system operation within our management.

\emph{Detected.} 
An uncertainty enters the lifecycle in the \textit{detected} state when it is first detected. 
Detection may result from incomplete, unstable, conflicting, or ambiguous information observed or reasoned during the system operation. 
At this stage, the uncertainty is recognized without yet establishing its characteristics.

\emph{Characterized.}
In the \textit{Characterized} state, the detected uncertainty is formally analyzed and represented. 
Its type, scope, severity, confidence, associated evidence and risk are determined and captured, which provides a structured basis for subsequent handling and decision-making. 

\emph{Mitigated.}
The \textit{Mitigated} state represents active uncertainty handling. 
Based on available evidence \(E(t)\), the framework applies mitigation actions, such as data acquisition, multi-agent reasoning, verification, or clarification, to reduce or bound the uncertainty. 
The uncertainty may remain in this state while mitigation continues or as new evidence is incorporated.

\emph{Resolved.}
An uncertainty transitions to the \textit{Resolved} state when accumulated evidence \(E(t)\) sufficiently reduces its severity and associated risk, allowing normal system operation to proceed without additional safeguards.

\emph{Expired.}
The \textit{Expired} state indicates that a decision has been taken despite residual uncertainty. 
Uncertainty is explicitly accepted and recorded, linking it to the committed decision to ensure traceability and accountability.

\emph{Escalated.}
Based on accumulated evidence \(E(t)\), uncertainty may transit from the \textit{Mitigated} state to the \textit{Escalated} state when associated risk remains high or cannot be adequately bounded through automated handling.
In such cases, uncertainty-handling actions \(H(u)\) transfer decision authority to higher-level agents or human operators for judgment, oversight, or governance.


\subsection{Uncertainty Representation}

\emph{How to represent uncertainty.}
We propose to represent uncertainty with the PSUM international standard~\cite{omg2024psum}, as it provides a formally defined semantic foundation for modeling uncertainty, evidence, and risk.
PSUM enables uncertainty to be expressed in a machine-interpretable and interoperable manner, making it suitable for coordination among multi-agent systems and deployment in critical domains such as healthcare.

In our framework, uncertainty is represented as a time-indexed PSUM-compliant object associated with all operational artifacts.
These artifacts include system inputs, intermediate reasoning artifacts, decisions, and executed actions, reflecting the fact that uncertainty may arise and evolve throughout the entire reasoning and execution process in multi-agent context. 
With PSUM, an uncertainty instance can be formally modeled as:
\[
u(t) = \langle 
\text{type}, 
\text{scope}, 
O(t), 
P(t), 
E(t), 
c(t), 
R(t), 
\tau,
U_{\uparrow}(t),
U_{\downarrow}(t)
\rangle
\]
where \textit{type} denotes an uncertainty category (e.g., data, model, inferential),
\textit{scope} defines the affected system components or decisions,
$O(t)$ captures associated ontological uncertainty that reflects irreducible variability or environmental indeterminacy,
$P(t)$ captures provenance and validity over time, 
$E(t)$ denotes observed evidence or reasoned info, 
$c(t)$ represents confidence or belief state, 
$R(t)$ denotes associated operational risk,
and $\tau$ defines temporal validity or expiration that governs state transitions in the uncertainty lifecycle.
To represent interdependencies and propagation among uncertainties in a multi-agent context,
$U_{\uparrow}(t)$ denotes the set of \emph{upstream uncertainties} on which the current uncertainty depends, such as uncertainties originating from upstream agents, data sources, or prior reasoning steps.
$U_{\downarrow}(t)$ denotes the set of \emph{downstream uncertainties} that are influenced or induced by the current uncertainty, capturing how uncertainty propagates across agents, tasks, and decision pipelines.

Uncertainty is inherently dynamic and evolving: as new evidence is observed, uncertainty may increase or decrease, prior assumptions may be invalidated, and associated risk may change.
By explicitly modeling uncertainty as a lifecycle-aware and time-indexed object, this representation enables uncertainty to be systematically queried, characterized, propagated, updated, and audited across its identification, characterization, mitigation, escalation, and resolution states during system operation.

\subsection{Uncertainty Identification}

\emph{How to detect and characterize uncertainty.}
Uncertainty is identified through continuous detection mechanisms operating across multiple layers of the system, together with representation mechanisms that characterize the identified uncertainty.
This process is primarily carried out by the \textit{Observer}, \textit{Reasoner}, and \textit{Constructor}.

The \textit{Observer} continuously perceives operational data from the environment and system state, while the \textit{Reasoner} analyzes observations and intermediate reasoning results to accumulate time-indexed evidence \(E(t)\).
Based on this evidence, the \textit{Constructor} generates uncertainty instances by detecting their presence and formally characterizes them using PSUM-compliant representations.

For example, at the data level, uncertainty is identified by the \textit{Observer} through validity checks,
completeness constraints, cross-source inconsistency detection, and distributional shift indicators.
At the reasoning level, the \textit{Reasoner} analyzes inference results produced by LLM-based agents, which are required to emit calibrated confidence estimates and explicit evidence links; divergence among independently reasoning agents, unsupported conclusions, or missing mandatory tool invocations trigger inferential or model uncertainty.
At the interaction level, ambiguity detectors, schema validation, concurrency conflicts, and feedback-loop patterns identify interpretational and interaction uncertainty.

In addition, higher-level operational indicators, such as frequent human overrides, repeated escalations, or persistent unresolved discrepancies, are treated as evidence of latent or systemic uncertainty.
All identified uncertainty is explicitly instantiated, recorded in the shared uncertainty registry, and explicitly linked to the affected agents, artifacts, and decisions to support subsequent evolution and adaptation.

\subsection{Uncertainty Evolution}

\emph{How uncertainty evolves based on accumulated evidence.}
Uncertainty evolution describes how uncertainty changes over time once it has been \textit{identified} and \textit{characterized}.
Rather than treating uncertainty as static, the framework models uncertainty as a dynamic object whose property may improve, persist, or deteriorate as new info becomes available.

The framework distinguishes epistemological uncertainty, which can often be reduced through additional evidence and actions, from ontological uncertainty, which cannot be eliminated and must instead be bounded and managed.
\textit{Reduction} of uncertainty is managed by two mechanisms, i.e., \textit{evolution} and \textit{adaptation}.
Evolution mechanisms are designed for reducing or refining uncertainty based on info gathering and evidential analysis that do not directly influence system behavior, while actions or strategies that modify system execution for the purpose of uncertainty reduction are handled by adaptation mechanisms.

Uncertainty evolution is driven by coordinated interactions among the \textit{Observer}, \textit{Reasoner}, and \textit{Evolver}.
The \textit{Observer} continuously monitors the environment and system state, collecting new observations and contextual info.
The \textit{Reasoner} analyzes these observations and intermediate reasoning results, integrating new info, reassessing prior assumptions, and updating the accumulated evidence \(E(t)\).
Note that, during the evolution phase, human actors exercising external authority may judge that an uncertainty can be resolved or explicitly expired based on their expertise.
Because such human interventions are neither routed nor orchestrated by the framework, they are incorporated solely as additional human-provided evidence \(E(t)\).
Based on the evolving evidence, the \textit{Evolver} revises the uncertainty state by updating its severity, confidence, and position within the uncertainty lifecycle.
Uncertainty evolution is modeled as an event-driven state transition:
\[
u(t+1) = f(u(t), e),
\]
where \(e\) represents events in which new evidence is acquired or becomes available, such as newly observed data, agent reasoning outputs, human-provided evidence, or time-based expiration.

Following characterization, uncertainty transitions from the \textit{Characterized} state to the \textit{Mitigated} state when evidential analysis motivates the initiation of uncertainty reduction or bounding activities.
As additional evidence accumulates, uncertainty in the \textit{Mitigated} state may transition to the \textit{Resolved} state when its severity and associated risk is acceptable (e.g., based on policy-defined thresholds), or to the \textit{Expired} state when a bounded decision is taken despite residual uncertainty.
However, uncertainty may transition from the \textit{Mitigated} state to the \textit{Escalated} state when accumulated evidence indicates an increase in uncertainty level or when the associated risk remains high.
Conversely, uncertainty in the \textit{Escalated} state may transition back to the \textit{Mitigated} state when additional evidence enables further refinement or reduction of the uncertainty.

All evolution steps are explicitly recorded in the uncertainty registry, ensuring traceability of how uncertainty changes over time and how evidence contributes to lifecycle transitions.

\subsection{Uncertainty Adaptation}

\emph{How to safeguard system operation via uncertainty-aware adaptation.}
In contrast to \textit{evolution}, which focuses on the continuous and explicit capture of uncertainty based on accumulated evidence, adaptation mechanisms are designed to safeguard system operation by regulating actions according to uncertainty severity and associated risk.
Adaptation governs how the system adjusts its behavior, autonomy, and decision pathways in the presence of evolving uncertainty.

Adaptation is realized by coordinating the \textit{Observer}, \textit{Reasoner}, \textit{Orchestrator}, and \textit{Commander}.
The \textit{Observer} and \textit{Reasoner} continuously provide updated evidence regarding system state, agent reasoning outcomes, and uncertainty evolution.
This evidence forms the basis for adaptive decision-making.

The \textit{Orchestrator} evaluates uncertainty states and accumulated evidence under policy and risk constraints, and determines which adaptive actions should be taken, including autonomy adjustment, verification requirements, workflow restructuring, or escalation.
The \textit{Commander} is responsible for executing the selected actions and enforcing system-level constraints, ensuring that adaptations are carried out in a controlled and auditable manner.
When necessary, the \textit{Orchestrator} may also notify relevant human operators and transfer decision authority to support judgment, oversight, and governance.

Adaptation operates in close coordination with planning, reasoning and coordination components in multi-agent systems and depends on the degree of autonomous authority granted to the framework.
The scope of adaptive actions is therefore determined by system policies that specify which aspects of execution the framework is permitted to modify.
For instance, at the planning phrase, when the framework is authorized to modify execution strategies, it may adapt plans by switching between sequential execution and collaborative multi-agent deliberation upon detecting inferential uncertainty or agent disagreement.
Planned actions may be deferred, decomposed into smaller reversible steps, or reformulated to explicitly surface unresolved uncertainty.
Regarding reasoning, adaptation regulates how agents produce and validate conclusions under uncertainty.
For example, the orchestrator may require additional evidence links, mandate tool-based verification, invoke specialist or critic agents, increase deliberation depth, or trigger parallel independent reasoning to quantify disagreement.
When uncertainty remains high, the orchestrator may constrain permissible inferences, request clarification, or escalate to human review.
For interactions among agents, adaptation addresses uncertainty arising from coordination and inter-agent dynamics rather than from individual agent reasoning.
When interaction uncertainty is detected, such as conflicting actions, incompatible assumptions, concurrency conflicts, or unstable feedback loops, the orchestrator adapts how agents interact, rather than what they reason about.
This includes enforcing coordination constraints, introducing arbitration mechanisms, limiting concurrency or execution rates, and restructuring interaction patterns to ensure consistent and stable collective behavior.
Through these mechanisms, adaptation prevents local uncertainties from propagating or amplifying across agents, thereby safeguarding system operation.

Human actors are actively involved in adaptation mechanisms, participating as decision-makers by reviewing uncertainty annotations, supporting evidence, and potential consequences, and by guiding or authorizing adaptive actions.
Human interventions are treated as first-class adaptation decisions.

\subsection{Uncertainty Management with HITL}

In complex, open-ended, and safety-critical environments (e.g., multi-agent systems), uncertainty may involve ambiguous intent, conflicting evidence, ethical considerations, or high-impact decisions that may not be handled properly by automated reasoning alone.
Therefore, we incorporate HITL into our uncertainty management framework, enabling human participation across multiple mechanisms, such as uncertainty evolution and adaptation, to support such as judgment, oversight, and accountable decision-making.
Our framework treats humans as \textbf{uncertainty-aware agents}, thereby ensuring robust, accountable, and risk-aware system operation under real-world uncertainty.

\emph{Human roles in uncertainty management.}
Humans play several roles in uncertainty management, with four primary roles contributing at distinct points in the uncertainty lifecycle:
(1) \textit{Interpretation:} resolving semantic ambiguity, intent uncertainty, and context-dependent meaning that automated agents cannot reliably disambiguate;
(2) \textit{Judgment:} weighing competing hypotheses when inferential uncertainty persists despite deliberation and verification;
(3) \textit{Risk acceptance:} explicitly accepting residual uncertainty for high-impact or irreversible decisions;
and (4) \textit{Governance:} providing accountability, ethical oversight, and compliance validation.
These roles align human involvement with specific uncertainty states and transitions, rather than ad hoc intervention.

\emph{Uncertainty-driven human engagement.}
Human involvement is not constant or regular, but triggered by \textit{policy-defined uncertainty concerns} within the orchestrator or as part of a business scenario.
Specifically, human engagement is initiated when uncertainty severity or associated risk exceeds the reliable scope of automated handling, such as in situations involving persistent disagreement among expert agents, unresolved high-severity data gaps, or decisions with significant safety, legal, or ethical implications.
This choice ensures that human expertise is applied only when necessary, while minimizing unnecessary intervention in low-risk or well-bounded decisions.

\emph{HMI for uncertainty-aware interaction.}
When engaging humans, the system is required to present uncertainty via a dedicated HMI.
The interface provides a structured view of the current decision or recommendation, the set of unresolved uncertainties, the supporting and conflicting evidence, the assumptions made by the system, and the potential consequences of action or inaction.
This presentation enables humans to resolve, contextualize, or explicitly accept uncertainty with relevance and awareness, instead of being asked to simply approve or reject outputs.

\emph{Human actions in uncertainty evolution and adaptation.}
In our framework, human inputs are treated as first-class events that directly participate in multiple uncertainty management mechanisms.
Human actions contribute to \textit{uncertainty evolution} by refining, reducing, or accepting uncertainty through clarification, expert judgment, or explicit risk acknowledgment, which are incorporated as human-provided evidence \(E(t)\).
They also contribute to \textit{uncertainty adaptation} by guiding or authorizing changes in system behavior, represented as uncertainty-handling actions \(H(u)\), when automated handling is insufficient.
In addition, human interventions may introduce new uncertainty that must be explicitly \textit{detected} and subsequently \textit{characterized} and managed by the framework.
All such actions are recorded in the uncertainty registry and linked to the affected uncertainty instances and decisions, preserving traceability and accountability across the entire uncertainty lifecycle.


\section{Conclusion and Future Work}
\label{sec:conclusions}

As LLM-based multi-agent systems are transitioning from experimental prototypes to mission-critical deployments like lifespan echocardiography, the ability to systematically manage uncertainty becomes a foundational requirement for clinical trust and software system reliability. This paper has moved beyond viewing uncertainty as a mere statistical error, reframing it as a dynamic lifecycle challenge inherent to the software engineering of autonomous systems. By establishing a rigorous understanding of epistemological and ontological uncertainties and proposing a framework grounded in uncertainty representation, identification, evolution, and adaptation, we provide a blueprint for systems that do not just experience uncertainty but actively reason about it and hence manage it systematically.

To ground this approach in established software standards, we examined the feasibility of applying the international standard Precise Semantics for Uncertainty Modeling (PSUM) as the formal foundation for 
representing these uncertainties. 
Within this paper, we aim to systematically examine, classify, and model all prevalent types of uncertainties of our primary industrial case study, 
and then propose a unified framework for managing each identified uncertainty type and associated components in a multi-agent setting within the case study: LLM-based echocardiographic software systems. 
Through this real-world application, we will discuss the practical challenges and demonstrate the feasibility of realizing our proposed framework, showing that a structured uncertainty lifecycle significantly enhances the transparency and calibration of diagnostic reasoning in high-stakes clinical environments.
%
In the future, we aim to implement and generalize this framework to other safety-critical domains. 



\bibliographystyle{ACM-Reference-Format} 

%



\newpage

\end{document}